# A Performance Analysis of Simple Runtime System for Actor Programming in C++

S.V. Vostokin, E.G. Skoryupina

Samara National Research University named after academician S.P. Korolev, Samara, Russia
easts@mail.ru

**Abstract.** In this paper, we propose the Templet – a runtime system for actor programming of high performance computing in C++. We provide a compact source code of the runtime system, which uses standard library of C++ 11 only. We demonstrate how it differs from the classic implementations of the actor model. The practical significance of the Templet was examined by comparative study on the performance of three applications: the reference code in C ++, managed by the OpenMP; the actor code in C ++, managed by the Templet; the actor code in Java, managed by the Akka. As a test problem we used a numerical algorithm for solving the heat equation.

## 1      Introduction

Actor model proposed by Hewitt in 1973 [1] isn't out of date; on the contrary, it attracts more and more attention to the developers. This is due to the modern trend of widespread hardware solutions for massively parallel computing applications. One of the main features of the actor model is the ability to describe an unbounded natural parallelism. Therefore, actively developing technologies such as the infrastructure software, Internet of Things and high performance computing, which uses massively parallel computations, fit well into the concept of actors [2].

There is a common framework Akka [3] in the area of infrastructure software and the Internet of things for interpreted Scala and Java languages. As for the high performance computing, where compiled languages dominate, actors have been little used. In our opinion, this is due to the prevailing stereotype of the implementation complexity of the actor model for compiled languages. New features of the latest versions of the standard, starting with C ++ 11, led to the development of effective and portable implementations of actor models for compiled C ++ [4].

The aim of the work is (1) to present a scalable, build-in implementation of the actor model in C++11, (2) demonstrate the effectiveness of the implementation by using high-performance computing test.

The remainder of this paper is organized as follows. First, we discuss the test problem for the performance evaluation from high performance computing. Then we de-scribe the implementation of the Templet actor runtime system. Further we propose three parallel implementation of the test problem: first based on the OpenMP, second - on the Templet system and the last - on the framework Akka. Here we describe the conditions of the computational experiment and compare the results. As a result, we conclude.

## 2    The Method of Efficiency Evaluation: Heat Equation Test

For comparative analysis we used an algorithm describing the solution to the heat equation (see Listing 1, 2). The algorithm was chosen due to the fact that it describes the sample implementation of frequently used finite-difference method and trivial one-dimensional decomposition of the data area for parallelization.

The constants W and H keep the width and height of the field grid area. The constant T is the number of time samples. An elementary operation op (Listing 1) shows the use of a differential stencil for calculating field values at the next time step.

```
1 void op(int i)
2 {
3   for (int j=1; j<W-1; j++)
4     field[i][j]=(field[i][j-1]+field[i][j+1]+
5       field[i-1][j]+field[i+1][j])*0.25;
6 }
```

**Listing 1.** Elementary operation of the heat equation benchmark

By changing the code of the elementary operation op (see Listing 1), we can derive algorithms for solving other problems. For example, it is easy to adapt an algorithm for three-dimensional field domain, without changing the overall calculations' structure (see Listing 2).

```
1 double seq_alg()
2 {
3   for (int t=1;t<=T;t++)
4   for (int i=1;i<H-1;i++) op(i);
5 }
```

**Listing 2.** Sequential algorithm for the heat equation benchmark

Another feature of the test is re-using the values of the temperature field in the calculation of the time layer by Seidel method. From an algorithmic point of view, this makes the relationship between iterations for i (Listing 2), which result in non-trivial issues to create a parallel solution based on the OpenMP.

## 3 The Templet Runtime System

The Templet actor computing system consists of three main parts: two primitive operations (send and access) and a function for working threads to process messages (tfunc). In the following listings we illustrate the mechanism for parallel execution of actors in the shared memory, using the C ++ standard library threads.

A sending message operation source code is shown in Listing 3. To send message is to place the message in a shared queue and notifying a thread, which may expect the message in the empty queue (line 6). The queue is protected by a mutex. It is captured in line 5. The message contains a reference to the actor-destination and the sign of being sent. They are initialized in line 4. Line 3 presents a guard of resending message. Resending message is an emergency situation, indicating that an error occurred in the application code.

```
1  inline void send(engine*e, message*m, actor*a)
2  {
3    if (m->sending) return;
4    m->sending = true; m->a = a;
5    std::unique_lock<std::mutex> lck(e->mtx);
6    e->ready.push(m);e->cv.notify_one();
7  }
```

**Listing 3.** Primitive operation 'send'

Access to the message object during the actor's processing procedure allowed, if the function access (see Listing 4) returns "true". Access granted if (1) the message refers to the actor, which is call the function; (2) the message is not on delivery (line 3). This condition is an invariant during the processing of a message in the context of a particular actor, until the message won't be sent by send operation. Sending messages is allowed only if the actor has an access to the message (see Listing 4).

```
1  inline bool access(message*m, actor*a)
2  {
3    return m->a == a && !m->sending;
4  }
```

**Listing 4.** Primitive operation 'access'

Source code of the working thread is shown in Listing 5. It implements a thread pool pattern [5]. The task in terms of the pattern is a message which is in the delivery. The message has a field sending value true. The sending field value of the message is true.

Polling the task from the queue (line 16), the working thread starts the actor's message processing procedure - recv. Preparation to the procedure recv includes several steps: (1) determination of the message's destination actor (line 19); (2) setting the lock on the actor (line 21); (3) changing the delivery sign of message on sending = false (line 22); activation of the recv procedure to process the message (line 23).

```
1  void tfunc(engine*e)
2  {
3   message*m; actor*a;
4
5   for (;;){
6     {
7       std::unique_lock<std::mutex> lck(e->mtx);
8       while (e->ready.empty()){
9         e->active--;
10        if (!e->active){
11                      e->cv.notify_one(); return;
12                }
13        e->cv.wait(lck);
14        e->active++;
15      }
16      m = e->ready.front();
17      e->ready.pop();
18    }
19    a = m->a;
20    {
21      std::unique_lock<std::mutex> lck(a->mtx);
22      m->sending = false;
23      a->recv(m, a);
24    }
25   }
26  }
```

**Listing 5.** Worker thread's function and 'recv' callback invocation

Note that the captured locks are released implicitly, when the thread leaves the syntactic scope of the object lock lck. Stop of the actor system computations occurs when there are no active working threads.

## 4    Parallel Algorithms for the Heat Equation Test

We implemented three parallel versions of the code in Listings 1, 2. All versions are driven by the following rules of parallelization. It is allowed to start iteration t along the time axis and i along the space axis (t, i), if (1) the iteration (t-1, i + 1) and (t, i-1) have completed; or (2), if t = 1 and iteration (t, i-1) has been completed. We assume that if iteration has no i + 1 or i-1 neighboring iteration, the neighboring iteration is completed. The first calculation iteration (1,1) is performed without a fulfillment of the conditions. The algorithm stops, when for each space coordinatei it is performed T iterations.

The considered computing algorithm can be implemented on the basis of the OpenMP, as shown in Listing 6. The idea of parallelism: either even or odd iterations i can be calculated simultaneously on each count t. A strict compliance with the rules of calculation guaranteed by the additional check in lines 5, 10, and 15 in Listing 6.

```
1 void par_omp()
2 {
3 #pragma omp parallel shared(H,T)
4 {
5 for (int t = 1; t <= (2 * T - 1) + (H - 3); t++){
6
7   if (t % 2 == 1){
8 #pragma omp for schedule(dynamic,1)
9     for (int i = 1; i < H - 1; i += 2)
10        if (i <= t && i > t - 2 * T) op(i);
11   }
12   if (t % 2 == 0){
13 #pragma omp for schedule(dynamic,1)
14     for (int i = 2; i < H - 1; i += 2)
15        if (i <= t && i > t - 2 * T) op(i);
16   }
17 }
18 }
19 }
```

**Listing 6.** OpenMP based parallel algorithm for the heat equation benchmark

Actors' implementations of the algorithm in listing 1, 2 allow to use the rules of parallelism explicitly. For this reason, each space coordinate i is matched by an actor. There are N = H-2 actors used in both actor algorithms.

In the the Templet implementation, the rules of parallelization presented in lines 5-7 of Listing 7. In lines 11 and 12, the actor informs its' neighbors i-1 and i + 1 (if any) of the completion of the iteration (t, i) by sending messages.

```
1 void recv(message* , actor* a)
2 {
3   int id = (int)(a - as);
4
5 if ((id == 0 || access(&ms[id - 1], a)) &&
6   (id == N - 1 || access(&ms[id], a)) &&
7   (ts[id] <= T)){
8
9   op(id+1); ts[id]++;
10
```

```
11    if (id != 0)     send(&e, &ms[id - 1], &as[id - 1]);
12    if (id != N - 1) send(&e, &ms[id], &as[id + 1]);
13  }
14 }
```

**Listing 7**. Actor based parallel algorithm for the heat equation benchmark, Templet runtime

In the Akka implementation, the rules of the parallelization are in lines 7-9 of Listing 8. In lines 13-20 the actor informs its' neighbors i-1 and i + 1 (if any) of the completion of the iteration (t, i) by sending messages. Note that the message handling code in Listings 7 and 8 is implemented identically for the convenience of comparison. The code block in lines 22-24 is for computations to stop.

```
1 public void onReceive(Object message) {
2      if (((Integer) message) == id - 1)
3              access_ms_id_minus_1 = true;
4      if (((Integer) message) == id)
5              access_ms_id = true;
6
7      if ((id == 0 || access_ms_id_minus_1) &&
8              (id == N - 1 || access_ms_id) &&
9              (Main.time[id] <= Main.T)) {
10
11         Main.op(id + 1); Main.ts[id]++;
12
13         if (id != 0) {
14                 Main.actors[id - 1].tell(id - 1, getSelf());
15                 access_ms_id_minus_1 = false;
16         }
17         if (id != Main.N - 1) {
18                 Main.actors[id + 1].tell(id, getSelf());
19                 access_ms_id = false;
20         }
21      }
22      if (Main.time[id] == Main.T + 1 && id == Main.N - 1) {
23              Main.system.terminate();
24      }
24 }
```

**Listing 8**. Actor based parallel algorithm for the heat equation benchmark, Akka

Both actor algorithms have an initialization code, which is not considered in the paper. Complete code of the Actor Templet library and the test cases are available at https://github.com/Templet-language/newtemplet/tree/master/etc/comparison_with_openmp.

## 5 Results

Computational experiments were performed on a computer with an Intel (R) Core (TM) i3-3220T RAM 4GB, Windows 10 x64. C ++ programs compiled in Microsoft Visual 2015. For Java programs we used the JDK version 1.8 and the library Akka version 2.4.17 deployed on the same computer.

The complexity of the problem is given by H. There are two space-time domain parameters of calculation: W = H * 2, T = H * 2 H, the both depends on H. Note that H also determines the granularity of computing. The bigger the parameter H is, the bigger chunks of data are processed sequentially.

Columns of Table 1 indicate the algorithms' duration time in seconds: $T_1^{JAVA}$ - a sequential Java implementation; $T_1^{NATIVE}$ - a sequential C++ implementation; $T_p^{AKKA}$ - a parallel Java implementation based on the Akka; $T_p^{TEMPLET}$ - a parallel C ++ implementation based on the Templet; $T_p^{OPENMP}$ - a parallel C ++ implementation based on the OpenMP.

To account for temporary fluctuations, the data presented in Table 1 has been statistically pre-processed. Each value in Table 1 is calculated by series of 19 experiments. The value includes only the significant digits, guaranteeing them from getting into the interval [min, max] with confidence factor of 90% (min - minimum, max - maximum time in a series of 19 experiments).

Table 1: Experimental computation time of the heat equation benchmarks

| H | $T_1^{JAVA}$, s | $T_1^{NATIVE}$, s | $T_p^{AKKA}$, s | $T_p^{TEMPLET}$, s | $T_p^{OPENMP}$, s |
|---|---|---|---|---|---|
| 400 | 1.79 | 1.357 | 0.8 | 0.42 | 0.40 |
| 500 | 3.5 | 3.028 | 1.2 | 0.92 | 0.9 |
| 600 | 6.1 | 5.238 | 1.8 | 1.81 | 1.77 |
| 700 | 9.8 | 7.37 | 2.7 | 3.01 | 2.92 |
| 800 | 14.6 | 12.46 | 3.7 | 4.60 | 4.52 |
| 900 | 20.9 | 17.73 | 5.4 | 6.5 | 6.4 |
| 1000 | 28.7 | 21.09 | 7.6 | 9.0 | 8.9 |

Table 2 shows the efficiency of the test implementation based on the proposed runtime system with a respect to the implementations based on the Akka and the OpenMP. The values EAKKA and EOPENMP show the percentage of the Templet implementation acceleration of reference implementations acceleration based on the Akka and the OpenMP.

Table 2: Relative efficiency of the Templet runtime system: $E_{AKKA} = T_p^{AKKA}/T_p^{TEMPLET}$ and $E_{OPENMP} = T_p^{OPENMP}/T_p^{TEMPLET}$

| H | $E_{AKKA}$, % | $E_{OPENMP}$, % |
|---|---|---|
| 400 | 190 | 95 |
| 500 | 130 | 99 |
| 600 | 99 | 98 |
| 700 | 90 | 97 |
| 800 | 80 | 98 |
| 900 | 83 | 98 |
| 1000 | 84 | 99 |

Correctness of the parallelization was checked by piecemeal test for equality of the temperature field values calculated by sequential and parallel method. We used equal random initial field values for parallel and sequential method. The physical interpretation of the calculation results was not carried out, since it is beyond the scope of this study.

## 6   Discussion

The experiments confirmed the high efficiency of the proposed simple implementation of the Templet runtime system for actor calculations. The Templet system has only a slight performance gap in tests performed by the OpenMP, and for the small H parameter's values 400..600 it is not far behind the Akka, or even exceeds it.

The advantage of actor algorithms for the Templet and the Akka is simplicity of implementation and debugging. The parallelism of the system is described in terms of a simple behavior of each individual actor. When using the OpenMP, it is required to understand the global state of computing at each time, resulting in complex boundary conditions of the algorithm cycles in Listing 6.

Our algorithm is not inferior to the implementation of the OpenMP. This result is obtained despite the fact that we used the expressive possibilities of C ++ Standard Library 11 to simplify the code, neglecting the efficiency. If necessary, it can be op-timized by using the primitive compare-and-swap, as proposed in [2], and by work stealing algorithms [6].

The source of the simplicity of our actor model implementation is a departure from the classical approach proposed by Agha [7] and implemented in the famous actor frameworks and languages, for example, Erlang [8], Scala [9], the CAF [2] and others. Agha's approach assumes that the messages are some values that are passed between the actors. They are accumulated in the mailbox - a special system structure associated with the actor. The actor has an access to the message values.

In our implementation, messages are treated as variables that store values. A programmer is not bounded to syntactic rules of access to the message from any actor at any time. However, an access is meaningful and does not lead to violations of logic provided that the function access for a couple of message-actor has returned true value. This approach does not require from runtime system the implementation of complex logic of copying values between the mailbox and the actor call frame.

The test also showed that despite the fact the native implementation of the test is superior to the implementation of Java performance, parallel implementation using the Akka is the best for the dimensions of the test problem for parameter H values 600 or more. This can be attributed to more sophisticated the Akka scheduling algorithm, than in the proposed Templet system, as well as the scheduling algorithm selected to test the implementation of OpenMP.

## Conclusion

We propose a simple implementation of the Actor model of computation in C ++ 11, and the possibility of its usage in high-performance computing. The test example of the heat equation illustrates the high effectiveness of the proposed implementation. It approximates to the effectiveness of the OpenMP, in some cases superior to the Akka and reduces the complexity of the coding.

Runtime library is used in object-oriented language Templet [10] for the implementation of parallel computing patterns. The patterns are used in solving problems of nonlinear dynamics in the design of spacecraft. [11] The authors express their gratitude to the Russian Foundation for Basic Research (RFBR) for partial support of these studies under the grant 15-08-05934 A.